\documentstyle[sprocl]{article}
\def\sla{\raise.15ex\hbox{$/$}\kern-.57em}

\def\Journal#1#2#3#4{{#1} {\bf #2}, #3 (#4)}

\def\NPB{{\em Nucl. Phys.} B}
\def\PLB{{\em Phys. Lett.}  B}
\def\PRL{\em Phys. Rev. Lett.}
\def\PRD{{\em Phys. Rev.} D}

\def\be{\begin{equation}}
\def\ee{\end{equation}}
\def\bea{\begin{eqnarray}}
\def\eea{\end{eqnarray}}
\begin{document}

\title{
\vspace{-5.0cm}
\begin{flushright}{\normalsize RUHN-00-5}\\
\end{flushright}
\vspace*{0.5cm}
REGULATED CHIRAL GAUGE THEORY
\footnote{Invited talk at the XVIII Autumn School
on Topology of Strongly Correlated Systems, 8-13 October, 2000, Lisbon}
}
\author{H. Neuberger}

\address{Department of Physics and Astronomy, Rutgers University, 
Piscataway,\\ NJ 08855, USA\\E-mail: neuberg@physics.rutgers.edu}


\maketitle\abstracts{ After a brief introduction to the overlap
two examples relating to topological properties of chiral fermion
systems in interaction with gauge fields are presented: It is shown
how the overlap preserves the continuum structure of exact fermionic
zero modes in gauge backgrounds that are instanton-like and
why chiral anomalies are inevitable. 
}
\section{Introduction}
On the lattice, any exact global symmetry group that acts locally
can be turned
into the gauge group of a gauge invariant theory by the lattice
version of the principle of minimal substitution and its non-abelian
generalization. This leaves no room for anomalies, and therefore
cannot produce the correct continuum limit. To get around this
difficulty one needs to take seriously the mathematical developments
of the mid eighties which taught us that the chiral determinant is
best thought of as a line bundle over gauge orbit space. This 
provides a natural setting for the nontrivial topological properties
that complicate the interaction of chiral fermions with gauge fields.
Also, it means that the fermionic path integral cannot be of an
entirely traditional kind, because if it were, the chiral
determinant would not stop at the stage of being a line bundle,
it would be just a function. The total disregard almost all workers
on lattice chirality had (and some still have) 
towards continuum topological properties explains
why a decade of efforts from the early eighties to the early nineties
amounted to an industry of failures. The key change in attitude 
that stared in 1992 and led to success was accepting the preservation 
of topological continuum properties on the lattice as essential. 
Although the Ginsparg-Wilson relation was in the literature since 1982,
the role of topology was not made explicit by it, and, as a result,
the GW relation only played an after-the-fact role in our understanding
of the modern solution to the chiral fermion problem. The main reason is that
the GW relation deals directly only with Dirac fermions, rather than 
the true fundamental building blocks of matter, namely the Weyl fermions. 

The plan for the rest of my talk is as follows: I shall first present
the main idea of the solution, make it concrete in the overlap and
proceed to explain what the two topological issues mentioned in the
abstract are and how they get resolved in the overlap. My main collaborator
in the overlap work was R. Narayanan. 

\section{Basic Idea}
It is easy (too easy!) to regulate in a gauge invariant way vector-like
gauge theories of generic structure:
\be
{\cal L}_\psi = \bar\psi{\sla D}\psi + \bar\psi \left ( P_L {\cal M} + 
P_R {\cal M}^\dagger \right ) \psi .
\ee
${\cal M}$ is the flavor-mixing mass matrix (taken as $N\times N$),
$\bar\psi,\psi$ are Dirac fermions and $D_\mu$ is the 
gauge-covariant derivative. Employing a bi-unitary fermion field
transformation, this theory is reduced to a theory
of $N$ decoupled flavors, each made out of two Weyl fermions. 

Consider now the case that $N=\infty$,
assuming that the above reduction
does not work because of the infinite dimensionality
of the flavor Hilbert space~\cite{fro,ovla}. Then, possibly, 
the number of Weyl fermions is infinite but odd, corresponding to
$\infty +{1\over 2}$ flavors. It is well known
how to get $\infty +{1\over 2}$ flavors: Pick
${\cal M}$ to have analytic index one.
\be
{\cal M} \psi =0 \Rightarrow \psi=\psi_0\ne 0,~~{\rm but}~~
{\cal M}^\dagger \psi =0 \Rightarrow \psi= 0.
\ee
Moreover, it is easy to find examples with ${\cal M}{\cal M}^\dagger$
strictly larger than zero, and in such a case the above structure
is stable under sufficiently small, but finite, perturbations
${\cal M}\rightarrow {\cal M}+\delta{\cal M}$. 

Obviously, a matrix and its hermitian conjugate can have different ranks
only for infinite dimension. All nonzero eigenstates 
of ${\cal M}^\dagger {\cal M}$ are paired with nonzero eigenstates
of ${\cal M}{\cal M}^\dagger$ and 
by rescaling ${\cal M}$ by a large
ultraviolet cutoff $\Lambda$ each pair becomes a Dirac fermion of
very large mass. Thus, only one light, zero mass Weyl particle is left
at low energies. 

This idea has some promise as a starting point because we start from
a bilinear fermion action with a formally gauge covariant kernel,
this kernel is for a vector-like structure and should therefore be easy
to regulate, and we avoid two deadly traps: The first is the ``shape trap''
and the second is the ``anomaly trap''~\cite{ovlb}.

Suppose that you try a Weyl-fermion action $\bar\psi_L K(A)\psi_L$,
where $K(A)$ is a gauge dependent finite matrix. When $A=0$ one
has the same number of $\bar\psi$ and $\psi$ fields and hence $K(A)$
is a square matrix. But, when $A$ is close to an instanton $K(A)$ must
become rectangular, say having one more column than rows. When
$A$ is close to an anti-instanton, we again need a rectangular
matrix, but now the number of rows exceeds that of columns by one. 
So, the shape of $K(A)$ must be allowed to change when the topological
properties of the gauge background change. Only an $\infty\times\infty$
matrix can effectively change its shape in the manner required.
If we start with a well defined finite dimensional
$K(A)$ we fall into the shape trap.

Gauge covariance means that under the replacement of $A$ by
a gauge transform $A^g$ 
\be
K(A^g ) = G(g)^\dagger K(A) G(g),
\ee
implying gauge invariance of $\det K(A)$ on account of the unitarity
of $G(g)$. Here we made no restriction on the representation
in which the fermions are and it could be that the continuum theory
is anomalous, which means that determinant is not gauge invariant.
It is possible to have a non-invariant determinant and a gauge covariant
kernel if the kernel is infinite dimensional and the definition of
the determinant is more subtle. Had we tried to keep $K(A)$ a finite
square matrix we would have fallen into the anomaly trap. 

Our task now is to make a concrete choice of ${\cal M}$, get rid
of the infinity of heavy Dirac particles and end up with something completely
well defined and finite, so it can be even put on a  computer. 

\section{Overlap}

We choose ${\cal M}$ to be an operator on integrable functions on the
real line, parameterized by $s$, 
\be
{\cal M}=\partial_s - f(s),
\ee
where $f(s)$ is $\Lambda$ for $s>0$ and $-\Lambda^\prime$ for $s<0$. This
arrangement means that 
\be
{\cal L}_\psi = \psi^* \partial_s \psi + \psi^* [\gamma_5 \sla D - \gamma_5
f(s)]\psi ,
\ee
where $\psi^* = \bar\psi\gamma_5$ plays the role of 
conjugate momentum to $\psi$ when $s$ is viewed as
an Euclidean time, and $\gamma_5 [\sla D - f(s)]$ is
a hermitian operator on account of the anti-hermiticity of $\sla D$
and its chiral property~\cite{ovla,ovlb,kap}. 
Assuming that $\sla D$ has been replaced by a finite
square matrix, the single infinity comes from the infinite extent of the
line $s$. The fermion path integral (at fixed gauge background)
is now immediately interpreted as giving
\be
e^{-E_{\rm g.s.}^- (A) \times \infty} \langle v_- (A)\vert v_+ (A)\rangle
e^{-E_{\rm g.s.}^+ (A) \times \infty}.
\ee
The infinite factors are gauge invariant, come from integrating out 
only very heavy Dirac particles, so are infinite but local gauge invariant
functionals of $A$, and therefore can be discarded leaving the regulated
chiral determinant given by
\be
\langle v_- (A)\vert v_+ (A)\rangle .
\ee
Here
\be
{\hat H}_\pm (A) \vert v_\pm (A)\rangle = E_{\rm g.s.}^\pm (A)v_\pm (A)\rangle
~~~~{\hat H}_\pm (A)={\hat a}^\dagger H_\pm (A){\hat a} ,
\ee
where
\be
H_+ (A) =\gamma_5 (\sla D + \Lambda)~~~~H_-=\gamma_5 (\sla D -\Lambda^\prime ).
\ee
The main point is that everything ended up being defined in terms of the finite
square matrices $H_\pm (A)$, and this is something a computer can 
``understand''.

A technical simplification can be made on the lattice: one can take the mass
parameter to infinity on one side of $s=0$ and consequently
replace the minus state in the overlap by a gauge field independent
reference state $\vert v_{\rm ref}\rangle$, defined by $H_-(A)=\gamma_5$.
We now turn to how the two topological issues mentioned in the abstract get
resolved, avoiding the shape and anomaly traps. 

\section{Instantons}
The operators ${\hat H}_\pm (A)$ conserve fermion number ${\hat N}_F$.
The reference state has fermion number equal to half of the dimension
of the matrix $H(A)\equiv H_+ (A)$, ${N\over 2}$. For any $A$, 
\be
{\hat N}_F \vert v(A) \rangle = n \vert v(A) \rangle .
\ee
For $A=0$ or nearby gauge fields $n={N\over 2}$ and the overlap
can be nonzero. But for an instanton and nearby gauge configurations
we have $n={N\over 2}+1$ and for an anti-instanton we have
$n={N\over 2}-1$. Thus, we get exact zero for the regulated chiral
determinants in topologically non-trivial backgrounds. Moreover,
the insertion of an ${\hat a}$ or an ${\hat a}^\dagger$ into the overlap
will render the result non-zero, exactly as required to generate
't Hooft vertices, once we interpret the insertions of ${\hat a}$
to correspond to insertions of $\psi$ in the path integral~\cite{ovlb}. 

We also ended up with a definition of lattice topological charge:
\be
Q_{\rm top} = n-{N\over 2}= -{1\over 2} Tr [ H (A)] .
\ee
The chiral fermion construction has naturally produced a slicing up of
the connected space of lattice gauge field configurations into regions
corresponding to different topological charge. A necessary ingredient
of any approach to the regularization of chiral fermions (which preserves
bilinearity) is a regulated definition of topological charge, 
something that was
missing from all of the previous failed attempts.

\section{Inevitability of anomalies}
 
We need first to see why the overlap is not guaranteed to be gauge
invariant although the Hamiltonian matrix is perfectly gauge covariant. Since $H(A^g ) = G^\dagger (g) H(A) G(g)$
and $H(A)$ is assumed to have no zero eigenstates, the ground state of ${\hat H(A)}$ is non-degenerate. Hence,
\be
\vert v(A^g )\rangle = e^{iS_{WZ} (g, A) } G^\dagger (g)
\vert v(A)\rangle .
\ee
Also, since $H_-=\gamma_5$, one can choose 
\be
G(g)\vert v_{\rm ref} \rangle =\vert v_{\rm ref} \rangle ,
\ee
implying
\be
\langle v_{\rm ref} \vert v(A^g )\rangle=
 e^{iS_{WZ} (g, A) } \langle v_{\rm ref} \vert v(A)\rangle,
\ee
which is a reflection of the fact that the ground state
is defined only up to phase. Thus, the overlap defines
a line bundle over the space of gauge orbits $\{A\}/\{{\cal G}\}$. 
This is exactly the required kind of mathematical structure ! 

One imagines starting from some reasonable smooth
\footnote{We shall discuss later
on what constitutes an acceptable initial phase choice.}
phase convention, a choice of a section in the line bundle
of $\vert v(A)\rangle$ over $\{A\}$.  One then considers
possible redefinitions of the phase by a local functional
$\Phi (A)$:
\be
\vert v(A)\rangle \rightarrow e^{i\Phi (A)} \vert v(A)\rangle .
\ee
The question of gauge invariance now amounts to whether
the Wess-Zumino functional is a trivial cocycle, meaning
that a $\Phi(A)$ can be found such that
\be
S_{WZ} (g, A)=\Phi(A^g )- \Phi (A) .
\ee
If such a $\Phi(A)$ can be found one can restore gauge
invariance. An anomaly occurs if such a $\Phi (A)$ does
not exist~\cite{geom}.

We now focus on the simple example of two dimensional $U(1)$ chiral
gauge theory and show that a topological obstruction makes
$S_{WZ}(g,A)$ a nontrivial cocycle. The obstruction should be
independent of the phase choice, invariant under the $U_0 (1)$
gauge group of phase redefinitions of $\vert v(A)\rangle$. The natural
$U_0 (1)$ invariant quantity is Berry's curvature ${\cal F}$, derived
from Berry's $U_0 (1)$ connection ${\cal A}$. Denoting $A_\mu (x) =\xi_\alpha$,
where $\alpha=(\mu,x)$, we have
\be
i{\cal A}=\langle v\vert \partial_\alpha v\rangle d\xi_\alpha
\ee
and
\be
i{\cal F}=id{\cal A}={1\over 2}[\langle \partial_\alpha v\vert \partial_\beta v\rangle
-\langle \partial_\beta v\vert \partial_\alpha v\rangle ] 
d\xi_\alpha d\xi_\beta .
\ee

Before continuing, we characterize the initial phase choice more precisely;
for our example we do not need a good section explicitly, only the
assumption that one exists.  
One of the main conditions the initial phase choice must obey is that
${\cal A}$ be a local functional of $A$. In addition, both perturbative
and nonperturbative anomalies must be reproduced, $S_{WZ}(g, A)$
must switch sign when the handedness of the fermion is switched and a
large discrete set of symmetries ought to be obeyed. At present,
the best phase choice theoretically seems 
to be the adiabatic phase choice~\cite{noncomp}
and the single numerically practical and theoretically 
plausible phase choice is the Brillouin-Wigner
one~\cite{ovlb}.

Because of $U_0(1)$ gauge invariance ${\cal F}$ is also a closed form
on orbit space $\{ A\}/\{{\cal G}\}$. Over $\{ A\}$, ${\cal F}$ is also
exact, but this is not guaranteed to be true over $\{ A\}/\{{\cal G}\}$.
However, if ${\cal A}$ were gauge invariant, ${\cal F}$ would be exact
also over $\{ A\}/\{{\cal G}\}$. If $S_{WZ} (g,A)$ could be eliminated
by a phase redefinition, Berry's connection would indeed be $U(1)$ gauge
invariant. We shall make the obstruction explicit by finding a two torus
embedded in orbit space over which the integral of ${\cal F}$ vanishes
only if ${\cal F}$ is made up additively of contributions of several
fermions of different handedness and charge, and the fermion set is
perturbatively anomaly free. When this condition is not met ${\cal F}$
is not exact and hence it is impossible to eliminate $S_{WZ}(g, A)$
by a $U_0 (1)$ gauge transformation.

We are working on a finite toroidal square 
lattice~\cite{geom} and the link variables
are $U_\mu (x)$. We pick a uniform background, $U_\mu (x) =e^{ih_\mu}$
for all links in the direction $\mu$. A shift $h_\mu\rightarrow h_\mu +
{{2\pi}\over L} n_\mu,~n_\mu\in Z$ amounts to a gauge transformation, so
in $\{ A\}/\{{\cal G}\}$ we have a torus $|h_\mu|\le {\pi\over L}$.
In Fourier space, $H(A)$ decouples into diagonal blocks $H_n$.
For $h_\mu =0$
\be
H_n=\pmatrix{{1\over 2} {\hat p}_n^2 -1 & i{\bar p}_n - {\bar p}_n^2\cr
-i{\bar p}_n - {\bar p}_n^2 & 1- {1\over 2} {\hat p}_n^2 },
\ee
where 
\be
{\bar p}_\mu = \sin p_\mu,~{\hat p}_\mu = 2\sin {1\over 2} p_\mu,~~~~
n_\mu=0,1,...L-1,~~~p_{n,\mu}={{2\pi}\over L} n_\mu .
\ee
For nonzero $h_\mu$, $p_n$ gets replaced by $p_n + h$. 

We need to calculate the two form
\be
f(h)=\langle {{\partial v}\over{\partial h_\mu}} \vert
{{\partial v}\over{\partial h_\nu}}\rangle dh_\mu dh_\nu .
\ee
$\vert v(h)\rangle$ is a Slater determinant of single particle two
component spinorial wave functions for each $n$, $u(p_n +h )$. Hence,
\be
f(h)= \sum_n \left [ {{\partial u^\dagger (p_n +h)}\over{\partial h_\mu}} \cdot
{{\partial u(p_n +h) }\over{\partial h_\nu}}\right ] dh_\mu dh_\nu ,
\ee
where
\be
H_n u(p_n) = E_n (p_n) u(p_n ) .
\ee
Although $u$ depends on phase choices, $f$ does not, and this is made
explicit by introducing the invariant two by two projector 
matrices $P=uu^\dagger$.
\be
f(h)={1\over 2}\sum_n Tr \left (P[\partial_{h_\mu} P, \partial_{h_\nu} P]
\right )|_{p_n + h}
 dh_\mu dh_\nu .
\ee
For each $n$,
\be
P|_{p_n+h} = {1\over 2} (1- {\vec w_n}(h)\cdot {\vec \sigma});~~~{\vec w_n}^2
(h)=1.
\ee
Hence,
\be
f(h)={i\over 2} \sum_n {\vec w_n}\cdot \left ( 
{{\partial {\vec w}_n}\over {\partial h_1}}\times
{{\partial {\vec w}_n}\over {\partial h_2}}\right ) dh_1 dh_2 .
\ee
One has ${\vec w}_n = {\vec w}(p_n +h)$ and ${\vec w}({\vec \theta })$
is a map from $T^2$ to $S^2$. We now calculate the integral
\be
\int_{|h_\mu|\le {\pi\over L}} f(h) .
\ee
The sum over $n$ combines with the $h$-integral to give the answer
\be
\int_{|\theta_\mu|\le {\pi}} {\vec w}(\theta )\cdot
\left ( {{\partial {\vec w} }\over{\partial \theta_1}}\times
{{\partial {\vec w} }\over{\partial \theta_2}}\right ) d^2\theta ,
\ee
which is the winding number of the map ${\vec w}(\theta )$, the integrand
having the explicit form of a surface element on the sphere, parameterized
by $\theta$. ${\vec w}(\theta)$ is defined by
\be
\pmatrix{{1\over 2} {\hat \theta}_n^2 -1 & i{\bar \theta} - {\bar \theta}^2\cr
-i{\bar \theta} - {\bar \theta}^2 & 1- {1\over 2} {\hat \theta}^2 }=
E(\theta ) {\vec w}(\theta ) \cdot {\vec \sigma};~~~{\vec w}^2(\theta )=1,
~~E(\theta ) <0. 
\ee

Until now we had only one left handed Weyl fermion of unit charge.
For charge $q$ each $h$-factor is multiplied by $q$ so we get
a multiplicative factor of $q^2$. The overall sign switches with
the handedness. Thus, unless $\sum q_L^2 = \sum q_R^2$,
the total ${\cal F}$ is proven not to be exact over gauge orbit space,
$S_{WZ}(g, A)$ cannot be eliminated, and no phase choice can restore
gauge invariance - the anomaly is inevitable. 

It is noteworthy that this inevitability was established in a completely
finite system, in the presence of both an IR and a UV cutoff; the main
ingredient in the proof was the smoothness in $h$ of the state entering
the overlap. Thus, the source of the obstruction is topological. 

\section{Discussion}

Let me end by touching upon some issues related to lattice
chirality that came up during the
discussion sessions at the school and pointing out connections to other
talks.

\subsection{How the Nielsen - Ninomiya obstruction in 
crystal momentum space is avoided}

Starting from the toroidal shape of momentum space on the lattice
one can establish under general assumptions that no traditional
free fermionic action can have truly chiral global symmetries, the
chiral nature being replaced by a vector-like one due to 
fermion doubling~\cite{nogo}. While this version of the no-go theorem holds
for any finite fermionic kernel, it can fail when the kernel
is an infinite matrix. In the case relevant to the overlap,
the free kernel in momentum space is an infinite operator analytically
depending on the momentum. However, the eigenvalues do no depend
analytically on the momentum: the mode corresponding to the chiral
fermion state exists only in a finite region around the origin of momentum
space and disappears outside it. Our first paper on the subject~\cite{ovla} was
devoted to showing that this abrupt behavior did not induce any
non-analyticity to any order in perturbation theory, since the 
fermionic propagator was well behaved.

In the vector-like case the shape trap does not apply because the
shapes of the kernels corresponding to the left and right handed 
blocks change in a complementary manner in the presence of instantons
or anti-instantons, so that the combined kernel in Dirac space is
always square. Thus, one should be able to get a traditional
type of action for vector-like fermions, so long the anomaly
trap is avoided. By a few manipulations
it was shown that the removal of all the extra heavy Dirac fermions
leaves a lattice action known as the 
overlap Dirac operator, $D_o$~\cite{do}.
Although the relation of $D_o$ to the GW relation was pointed out
already in~\cite{dw} this comment seems to have been ignored until it
was amplified in~\cite{dogw}. $D_o$ has no difficulty with the no-go theorem
because it does not anticommute with $\gamma_5$; however, the chiral
symmetry is just hidden as a result of eliminating all the extra fermions
and this can be re-interpreted as a consequence of the GW relation. 

\subsection{Numerical prospects}

It is premature to predict when the new fermionic actions would replace
the traditional ones, because at the moment implementation costs 
run roughly a factor of 100 higher. The most tested approach is
that of domain wall fermions~\cite{vranas} and the next one in line is based
on rational approximants to $D_o$~\cite{prac}. Personally, I believe that
using $D_o$ is better because it is cleaner theoretically and,
apparently, the implementation cost is similar~\cite{alt}. 
If the factor of 100
were not critical there would be no question that one should use the
new fermionic actions. This factor probably can be reduced by several tricks
and the last word hasn't been said yet. 

Already now, for some calculations, the simplification afforded
by exact chirality on the lattice may outweigh the large implementation
cost. This would be the case in calculations that deal with
matrix elements of operators that do not mix under renormalization
in the continuum but, on the lattice, do mix when traditional fermions
are employed. Disentangling this mixing numerically is so costly that
it is probably advantageous to pay the higher implementation price 
instead and use fermions with exact chiral symmetry.

\subsection{Restoring gauge invariance}

We have seen an example which shows how continuum anomalies prohibit the
restoration of non-perturbative gauge invariance. In the abelian case,
at infinite volume and with non-compact gauge fields one can show
that non-perturbative gauge invariance can be restored if 
continuum anomalies cancel~\cite{noncomp}.
From the mathematical viewpoint, since four dimensional abelian gauge
theories do not have an interacting continuum limit (just like $\phi^4$,
which is relevant to Higgs physics~\cite{higgs}), 
the difference between an anomalous and a non-anomalous theory is of
a more quantitative than qualitative nature. Both theories are
effective in the sense that they make predictions of limited accuracy
whose validity holds only for energies below a cutoff $\Lambda$.
In addition, the coupling constant cannot be too large, and has to
vanish in the limit $\Lambda\to\infty$. The bound on the coupling
constant is much less stringent if anomalies cancel~\cite{noncomp}.

\subsection{Connections to other talks}

J. Zinn-Justin has given an introduction to the GW relation in his
series of talks on the regularization of chiral gauge theories. The
overlap produces the overlap Dirac operator, which satisfies the GW
relation and in a certain sense is the most general solution to this
relation. It is important to keep in mind that satisfying the GW relation
is, in itself, not sufficient to ensure the presence of chiral fermions.
Extra conditions must be attached, and it is easy to come up with
useless solutions to the GW relation if these extra conditions are
not met. At present there is some discussion in the literature
about what the minimal set of extra conditions needed to ensure
the presence of chiral fermions is~\cite{chiu}.

The technical aspects of my derivation of the two dimensional
obstruction 
to the restoration of gauge invariance 
(showing how the ``anomaly trap'' is avoided)
are almost
identical to the derivations surrounding the TAP integers that
appeared briefly in the lectures by D. Thouless. Berry's
connection plays a central role, similar to the role it plays in 
the chiral fermion context. 

The ``shape trap'' is
associated with 't Hooft vertices which appeared in G. 't Hooft's lectures
in the vector-like context. The role of the 't Hooft vertex is
even more dramatic in the chiral context, as discovered by 't Hooft
%~\cite{baryon} 
but not discussed in his lectures here. 

If one keeps the extra fermions in the picture, and uses the 
entire system to calculate anomalies the quantization of the anomaly
coefficient comes from nontrivial winding in Fourier space
%~\cite{so}. 
This
makes contact with Volovik's lectures where the role of topology
in momentum-energy space (round singularities in the fermion
propagator) was emphasized.

\section*{Acknowledgments}
This research was 
supported in part by the DOE under grant \#
DE-FG05-96ER40559. I wish to thank the organizers 
for the invitation to participate, for support, and for the immense
hospitality extended.

\end{document}